\documentclass[preprint,12pt]{elsarticle}

\usepackage{graphicx}
\usepackage{amssymb}

\usepackage{lineno}

\newcommand{\rsun}{$R_{\odot}$}
\newcommand{\msun}{$M_{\odot}$}

\newcommand{\var}{\hbox{V393\,Scorpii~}}

\newcommand{\op}{\hbox{$\Phi_{o}$ =~}}

\journal{Pub. of the Astronomical Society of the Pacific}

\begin{document}

\begin{frontmatter}


\title{Evidence of active regions in the donor of the Algol-type binary \var and test for the  dynamo model of its long cycle.}



\author[1]{Ronald E. Mennickent}
\ead{rmennick@udec.cl}
\author[1]{Dominik R.G. Schleicher}
\author[1]{Rub\'en San Martin-Perez}
\address[1]{Departamento de Astronom\'{\i}a, Casilla 160-C, Universidad de Concepci\'on, Chile}


\begin{abstract}
V393\,Scorpii is a member of the subclass of Algols dubbed Double Periodic Variables (DPVs). These are semidetached binaries with B-type primaries showing a long-photometric cycle lasting in average 33 times the orbital period.  We describe the behavior of unreported metallic emission lines in the cool stellar component of this system. The emissions can be single or double for a same line and sometimes show velocity shifts regarding the velocity of the center of mass of the star. In addition, these lines are stronger during the high state.
This behavior suggests the presence of active regions in the surface of the rapidly rotating A7 donor covering a fraction of the visible hemisphere, which have larger emissivity during the high state.
Our finding supports the recently proposed dynamo model for the long cycle of DPVs proposed by Schleicher \& Mennickent.  The model predicts an increase of the dynamo number of the donor during epochs of mass transfer in this system, and a theoretical long/orbital period ratio very close to the observed one at the present system age.  
\end{abstract}


\end{frontmatter}


\section{Introduction}
\label{S:1}

Double Periodic Variables (DPVs)  are  a class of  Algol  binaries showing a long photometric cycle lasting about 33 times the orbital period; they are usually hotter and more massive than classical Algols and more than 250 of them have been found in the Galaxy and the Magellanic Clouds \citep{2003A&A...399L..47M, 2008MNRAS.389.1605M, 2010AcA....60..179P,  2012MNRAS.421..862M, 2012MNRAS.427..607M, 2013MNRAS.428.1594G, 2013A&A...552A..63B, 2013AcA....63..323P, 2014A&A...567A.140B, 2016MNRAS.461.1674M, 2017SerAJ.194....1M}.  Recently, \citet{2016MNRAS.455.1728M} showed, based on a limited number of well studied systems,  that DPVs are those semi-detached Algols with B-type primaries of radius slightly larger than the critical radius\footnote{The critical radius just allows the stream coming from the secondary  to hit the primary. } and also that they host relatively stable optically thick accretion discs of typical extension of a few tens of solar radii. Evolutionary studies of DPVs suggest that they are found inside or close to an episode of mass transfer due to Roche lobe overflow by the donor \citep{2016MNRAS.455.1728M}.

It has been suggested that a magnetic dynamo is responsible for the long cycles of DPVs \citep{2017A&A...602A.109S}. The idea is that the synchronized  donor rotates at a significant fraction of the critical velocity, and becomes unstable to 
inner convective motions, generating a magnetic dynamo similar to those observed in sun-like stars or possibly RS\,CVn binaries. The above authors shows that the dynamo period is proportional to the 
stellar rotational period and only weakly dependent of the stellar mass.  
Relevant to the discussion is the existence of chromospheric emission in the mass-losing star of some Algols, revealed in the emission cores of the Ca\,II H and K lines or the alternate changes of the orbital period  \citep{1989SSRv...50..311B, 1989SSRv...50..219H} or even in emissivity maps obtained with the technique of Doppler imaging \citep{2014ApJ...795..160R}.
 Chromospherically active regions have also been invoked as the cause for the coronal X-ray emission observed in some short-period Algol binaries by satellites as ASCA and ROSAT \citep{1993ApJS...88..199R, 1995ApJ...445..840S}.
 
In this paper we give support for the dynamo model of the long cycle of DPVs, based on a study of  spectroscopic data densely covering the orbital and long cycles of the DPV V\,393\,Scorpii. This binary has an  orbital period  of 7.71 days and  a  long cycle  of 253 days. 
 The basic stellar and orbital parameters have been obtained from the model of the light curve by \citet{2012MNRAS.421..862M}.
The system consists of a 2.0 \msun\ inflated donor transferring mass onto a 7.8 \msun\ gainer which is surrounded by an optically thick 
disc of radial extension  about 9.5 \rsun\ and vertical thickness about 2 \rsun. The orbital separation for the system is 35.1 \rsun\ and the system is observed under inclination of 80 degree.
The system shows enhanced Balmer emission during the long-cycle maximum attributed to a disc-wind and also C\,I\,6588 and Mg\,II\, 4481 emissions arising  from the donor,  attributing a chromospherically-active character to this star \citep{2012MNRAS.427..607M}.

While \citet{2012MNRAS.427..607M} presented the analysis of the main  optical spectroscopic features of V\,393\,Scorpii (i.e. Balmer and He\,I lines), here we present a study of  weak  metallic lines most of them not previously reported  and  show that these lines can be used as diagnostics of  the chromospherically active donor
star. In past studies, variable Balmer emission was interpreted as a signature of a bipolar wind emerging from the stream/disc impact region \citep{2012MNRAS.427..607M}. In addition, 
the study of H\,I infrared lines also provides evidence for the presence of an intermediate-latitude wind and suggests that material 
escapes in the orbital plane from the disk through the $L_3$ point \citep{2010MNRAS.405.1947M}. In order to place our new study into context, a summary of the spectral lines previously studied in this system, and its significance, is given in Table\,1.

 The finding of magnetic activity in DPV donors might have wide astrophysical significance, since, contrary to G/K/M dwarfs with solar-type 
magnetic cycles,  these relatively hotter A/F-type giants have not been much explored as sites of magnetic activity, much less as hosts of magnetic dynamos; in some cases magnetic activity is expected to be limited to the stellar core of A-type dwarfs, as result of  a fossil field  which might explain the Ap stars \citep{2009ApJ...705.1000F}.
It is possible that rapid rotation plus structural changes happening during expansion to the giant stage might favor the appearance of magnetic instability
behind the stellar surface bringing the dynamo into existence.

\begin{table}
\centering
 \caption{Spectral lines studies in V\,393\,Sco. Wavelengths are in nanometers.}
 \begin{tabular}{@{}lr@{}}
 \\ \hline
Reference \& &  Highlights\\ 
Line & \\
\hline
\citet{2010MNRAS.405.1947M} &Evidence of high-latitude wind \\
UV + infrared range  & and equatorial outflow. \\ 
\hline 
L$\alpha$ &  \\
Si\,IV 139.37, 140.28 & \\
Al\,III 185.47, 186.28 &  \\
Mg\,I 1081.4 &  \\
He\,I 1083.3, 1091.6 & \\
Si\,I 1084.7, 1088.8 & \\
Mg\,II 1091.7 & \\
\hline
\citet{2012MNRAS.427..607M} &Evidence of bipolar wind linked  \\
Optical range (stronger lines)& to the long cycle and active donor. \\
\hline 
H$\alpha$, H$\beta$, H$\gamma$ & \\
Mg\,II 448.1 & \\
C\,I 493.2, 658.8 & \\
He\,I 587.5, 667.8 & \\
Si\,I 634.7, 637.1 & \\
O\,I 777.3 & \\
\hline
This paper & Donor activity related to the long cycle.\\
Optical range (weaker lines)& \\
\hline 
Ti\,II 446.8, 4501.3&\\
Mg\,II 448.2&\\
Fe\,II 448.9, 450.8, 451.5, 495.8& \\
C\,I 493.2, C\,I 658.8 & \\
\hline 
\end{tabular}
\end{table}

This paper is organized as follow: in Section 2 we present a summary of our observations and methodology. In Section 3 we present the results of our spectroscopic analysis of Fe\,I, Fe\,II, Mg\,II, C\,I and Ti\,II lines. 
In Section 4 we discuss the origin of these lines supporting the existence of active regions in the donor star. In  Section 5 we explore the evolution of the binary along with the dynamo properties following  the Schleicher \& Mennickent model for the long cycle. Finally, in Section 6, we summarize the main conclusions  of this research.

\section{Observations and methodology}
\label{S:2}

The full description of the observational material, the observing log and details of data reduction and analysis can be found in  \citet{2012MNRAS.427..607M}.
Here we give only a brief summary.
The spectroscopic data span 3 years between 2007 to 2010  and consist of  555 spectra obtained with several 
spectrographs, mostly with resolution  $R \sim$ 40.000 and covering the wavelength region of 3350-9500 \AA, with S/N usually above 100.
They are corrected by earth translational motion and normalized to the continuum. The RVs presented in this paper are heliocentric ones,  they are obtained after subtracting the earth velocity relative to the Sun.  
In this paper we present the continuum normalized wavelength calibrated spectra and also the residual spectra obtained after subtracting the
theoretical donor spectrum, considering its fractional light contribution in each wavelength range and each orbital phase.
The ephemerides have the usual meaning,  phase $\Phi_o$ = 0.0 for the orbital cycle refers to the  primary eclipse, i.e. inferior conjunction of the donor, and phase $\Phi_l$ = 0.0 for the long cycle refers to the epoch of maximum brightness for this cycle.

\section{Results}
\label{S:3}

\subsection{Detection of metallic emission lines}

We find emission lines of the following elements Fe\,I, Fe\,II, Ti\,II, C\,I  and Mg\,II.
Weak emission, sometimes  double and sometimes single, were detected in the cores of  metallic lines even before processing the spectra, mainly near main and secondary eclipse. Specially notable 
is the double emission of the Mg\,II\,4481 line (Fig.\,1). The above lines are usually found in absorption in the atmospheres of A-type stars, but the character of emission indicates a different formation mechanism.  The presence of double emission in the context of close interacting binaries is usually interpreted in terms of rotation of the emitting material, and explained as the Doppler splitting of frequencies of photons coming from the receding and approaching sides of the emitting region. Double emission lines are usually interpreted as signatures of an emitting disk.

\subsection{Radial velocity and emission line strength variability }

The C\,I\,4932 emission is almost always seen as a single peak with a red absorption component possibly  Ba\,II\,4934.09.  
The radial velocity (RV) of C\,I\,4932 emission follows almost perfectly the donor RV (Fig.\,2).

The Ti\,II 4501.3, Fe\,II 4508.3 and Fe\,II 4515.3 single emissions follow a singular RV curve  with  $\gamma$ velocity close to the systemic velocity, viz.\,-12 km/s (Fig.\,2). At this and following graphs we show for comparison the theoretical RV curves of hypothetical emitting material co-rotating with the binary  at the inner Lagrange point ($L_{1}$) and the points located $\pm$ $R_d$ from the donor center along the orbital semi-major axis, here named  ``donor outer face''
and ``donor inner face'', respectively. 
For a semidetached system the point on the donor surface closer to the center of mass should be placed at  $L_{1}$, i.e. if the star fills its Roche lobe it departs from spherical symmetry. 
We notice that near \op  0.6 two distinct velocities for the single emission can be found. The RVs follow different reference lines suggesting the existence of active regions at different sides of the donor star.

At the red emission wing of the H$\alpha$ line we found C\,I\,6587.75 in emission. This can be found as single peak or double peak emission with a mean separation of 53 $\pm$ 9 (std) km  s$^{-1}$ (42 measurements averaged). Their radial velocity follows the donor RV. 
In all cases when emission is detected, this is single and never double at \op 0.5. On the contrary, at \op 0.0 both kinds of profiles are observed (Fig.\,3). This tendency is observed in all lines described in this section, 
Mg\,II 4482 (Fig.\,4) and Fe\,I\,4957.6 (Fig.\,5). The average peak separation for 14 double lines of Fe\,I\,4957.6  is 78 $\pm$ 7 km s$^{-1}$. Note that the single emission is not always observed with the donor RV, but sometimes with larger or smaller velocity,  appearing as a blue or red emission peak. The peak separations above are close to the projected rotational velocity of the donor, viz.\, 60 km s$^{-1}$
 \citep{2012MNRAS.427..607M}. This fact, along with the alternating single/double nature of the peaks, might indicate a rotational origin for the splitting of the lines, if formed in different active regions in the stellar surface.

The equivalent width of the C\,I\,4932 residual emission increases sharply near \op 0.0 and decreases smoothly at \op 0.5. The Mg\,II\,4482 residual emission also increases at  primary eclipse, but does not decrease at secondary eclipse.
It also increases around orbital phase 0.7 (Fig.\,6), something also observed in C\,I\,4932 but not so prominently.

\begin{figure}
\scalebox{1}[1]{\includegraphics[angle=0,width=7cm]{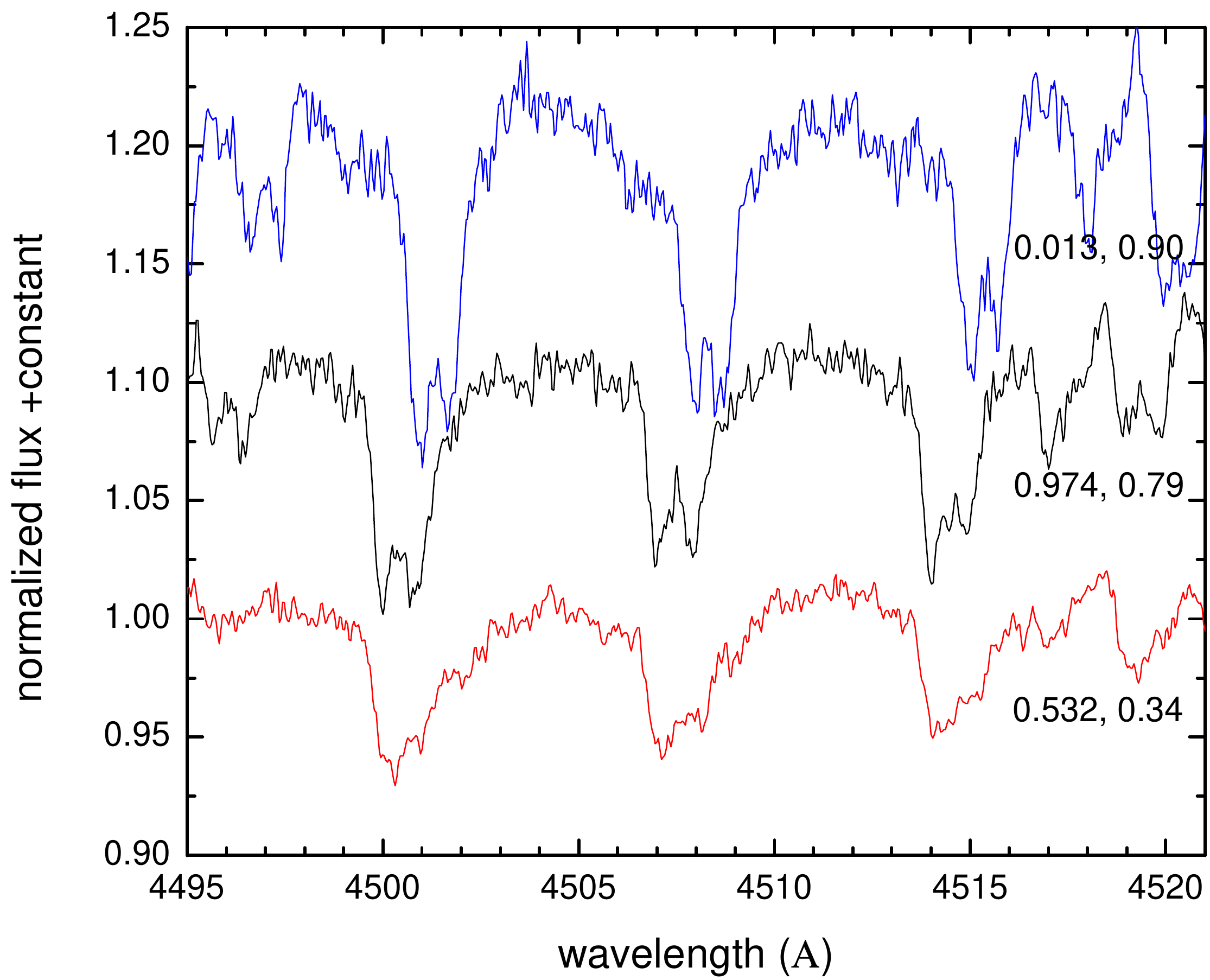}}
\scalebox{1}[1]{\includegraphics[angle=0,width=7cm]{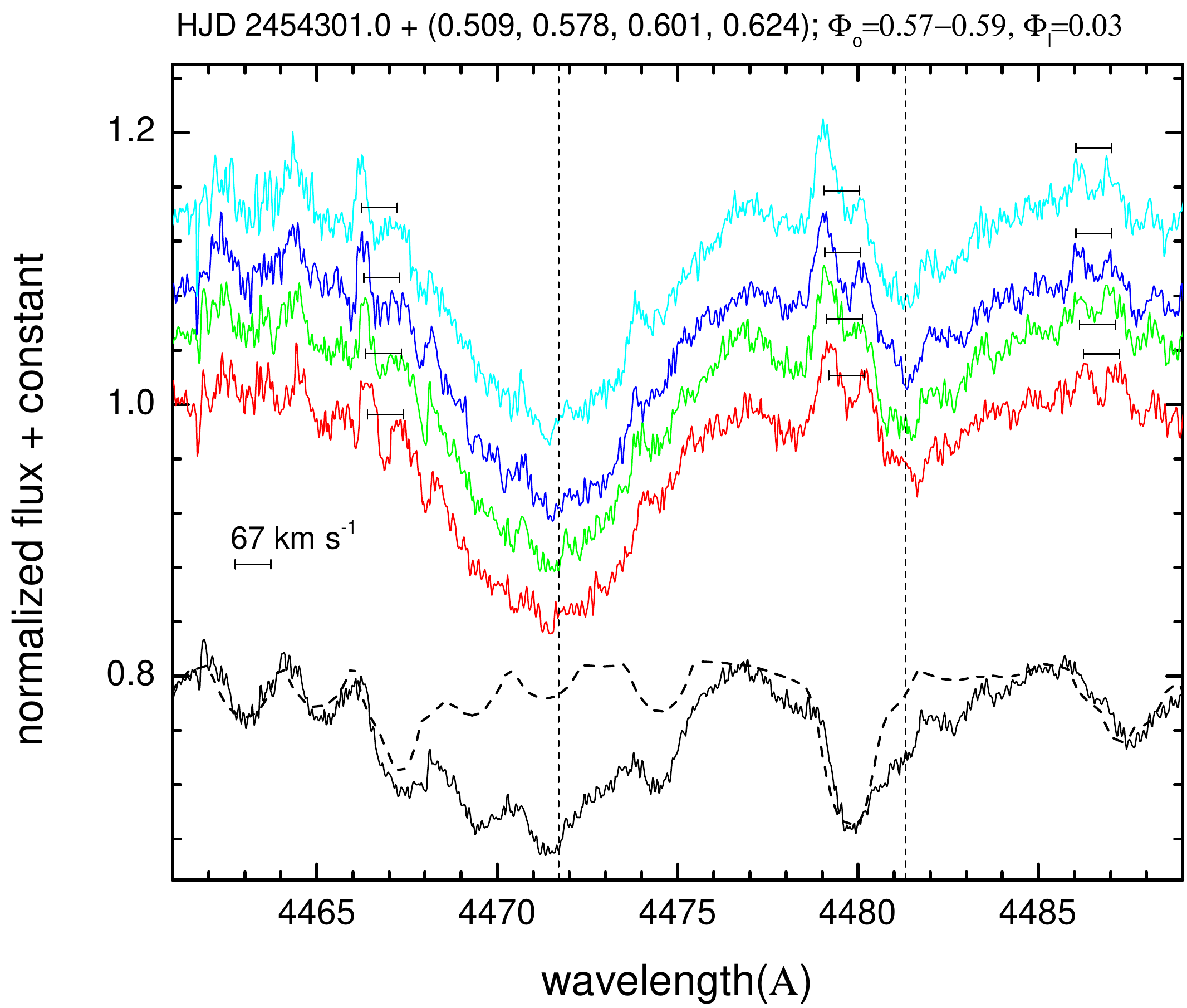}}
\caption{Left: Examples of core emission in the Ti\,II 4501.3, Fe\,II 4508.3 and Fe\,II 4515.3 lines. These spectra are not donor-subtracted and show that the detected emission is not an artifact of the process of removal of the 
underlying donor spectrum.
Labels  show the orbital and long-cycle phases. Right: Examples of double emission in Ti\,II\,4468.5, Mg\,II\,4481 and Fe\,II\,4489.2. Four donor subtracted consecutive spectra are shown (the last one at the top) along with marks indicating double emission profiles. As illustration of the effect of the subtraction procedure, the original  spectrum at HJD 2454301.578  (black solid line)  is over-plotted with its donor template (dashed line). Vertical dashed lines show the central wavelengths of  He\,I\,4471 and  Mg\,II\,4481.}
  \label{x}  
\end{figure}

\begin{figure}
\scalebox{1}[1]{\includegraphics[angle=0,width=7cm]{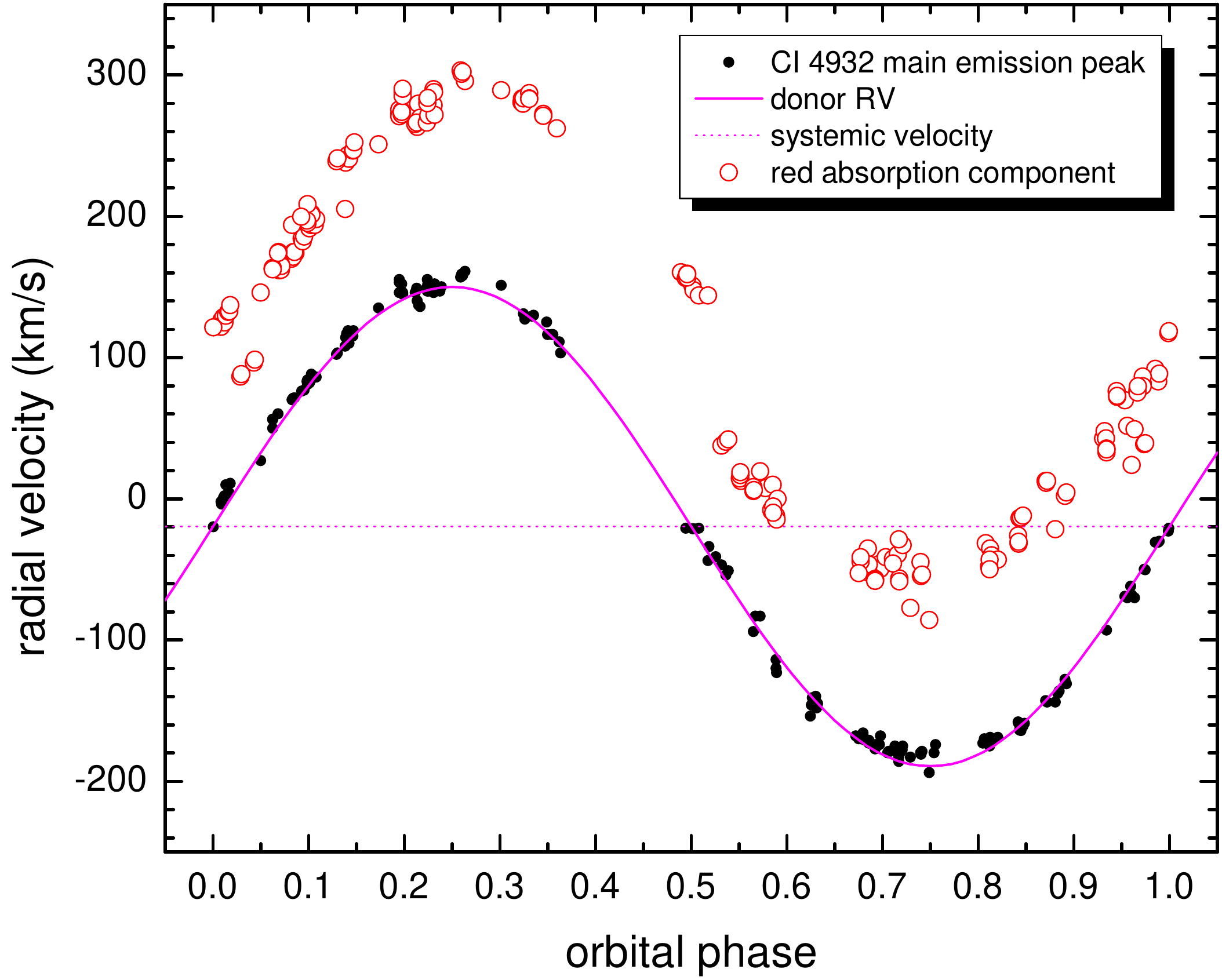}}
\scalebox{1}[1]{\includegraphics[angle=0,width=7cm]{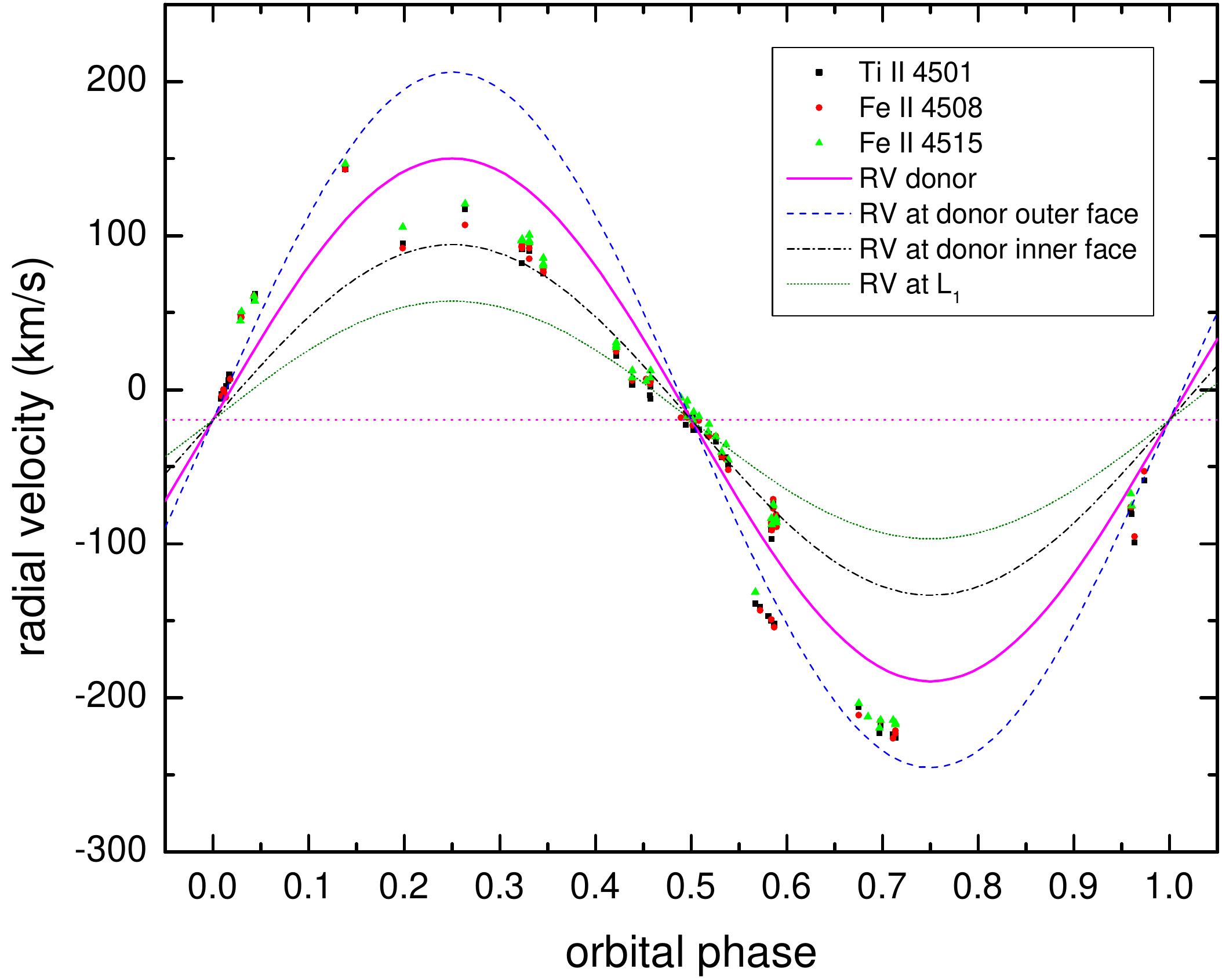}}
\caption{Left: The RV for the C\,I\,4932 main emission feature and its redward absorption line (possibly Ba\,II\,4934).  Right: The radial velocity for the single emission detected in the donor subtracted lines Ti\,II 4501.3, Fe\,II 4508.3 and Fe\,II 4515.3  along with synthetic RV curves.}
  \label{x}  
\end{figure}

\section{Origin of the metallic emission lines}
\label{S:4}

The analysis of combined RV and EW data allow us to place the origin of the metallic  emission lines at the donor star.
In fact, the increase
of the emission during  primary eclipse can be explained as non-occultation of the source  (the donor,  Fig.\,6) while the main continuum flux source (gainer)
is occulted. The decrease of the emission during secondary eclipse in CI\,4932 can be explained as occultation of the emitting source (the donor, Fig.\,6 right panel). This origin is also consistent with the emission radial velocity roughly following the velocity of the donor center of mass. The different behavior of the Mg\,II emission strength during the secondary minimum  (Fig.\,6 left panel) might indicate a slightly different place of formation, but still at the secondary star, as evident from its RV curve.

We notice that the fact that the emission equivalent width is similar at low and high stage (even larger in high stage  in occasions), indicates that the
relative contribution of the line emission remains almost constant. Since the continuum is larger during high stage this means that the intrinsic emission is larger at high stage.  

The peculiar positions of single or double emissions regarding the donor radial velocity can be interpreted in terms of active emitting regions in the chromosphere of the donor, which appear shifted in velocity 
according to their position in the atmosphere and its projected rotational velocity regarding the observer.  To establish the exact geometry of the active regions is beyond the hope of this paper, and probably far from the possibilities with the current data. A detailed analysis of the donor emission lines might, through the technique of Doppler imaging
or eclipse mapping, yield more light on this point.   

\begin{figure}
\scalebox{1}[1]{\includegraphics[angle=0,width=7cm]{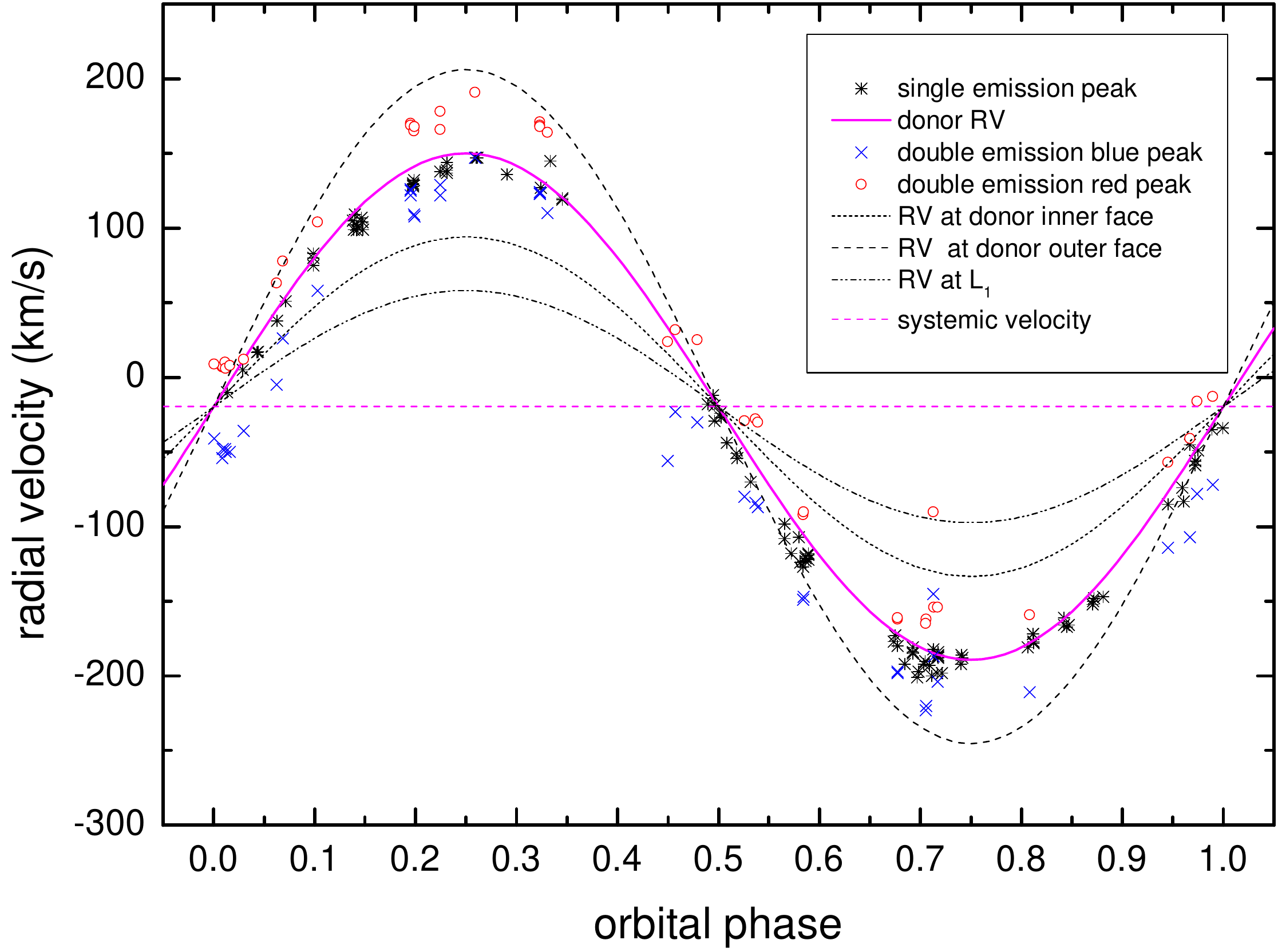}}
\scalebox{1}[1]{\includegraphics[angle=0,width=7cm]{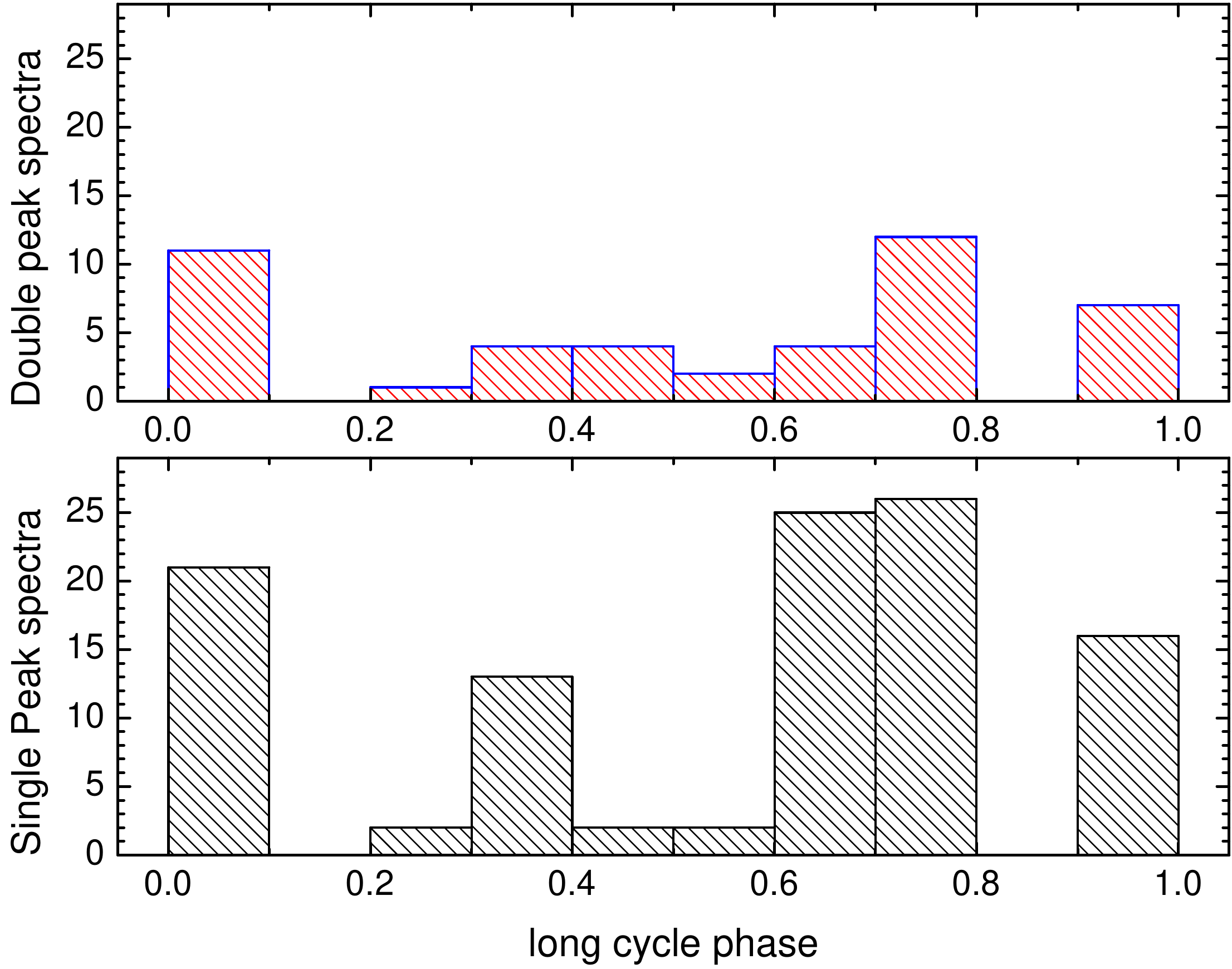}}
\caption{Left: Radial velocities for the C\,I\,6587.75 line. Some illustrative  synthetic RV curves are also shown. Right: Histograms showing the occurrence of single and double peak emission. } 
  \label{x}
\end{figure}

\begin{figure}
\scalebox{1}[1]{\includegraphics[angle=0,width=7cm]{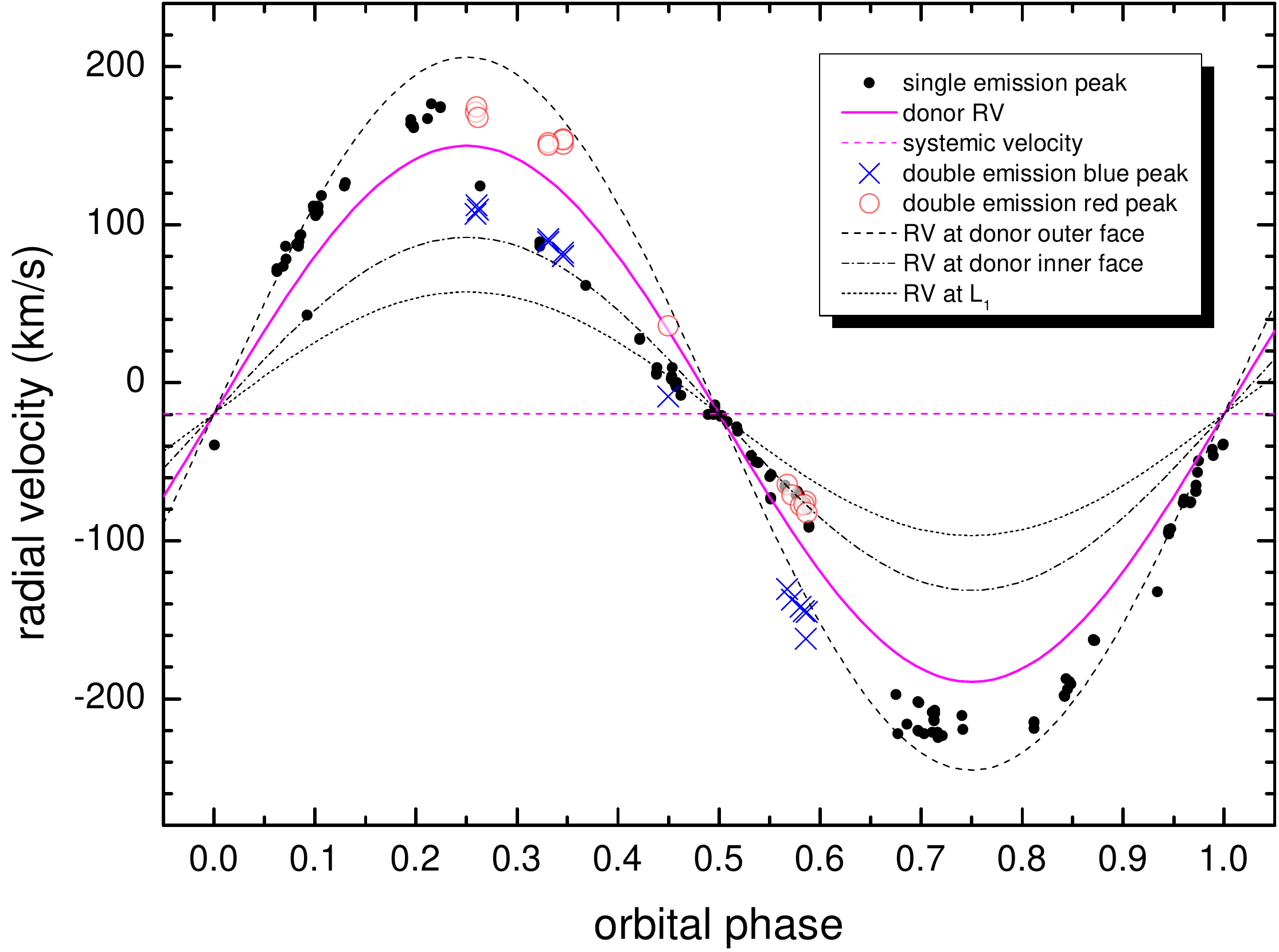}}
\scalebox{1}[1]{\includegraphics[angle=0,width=7cm]{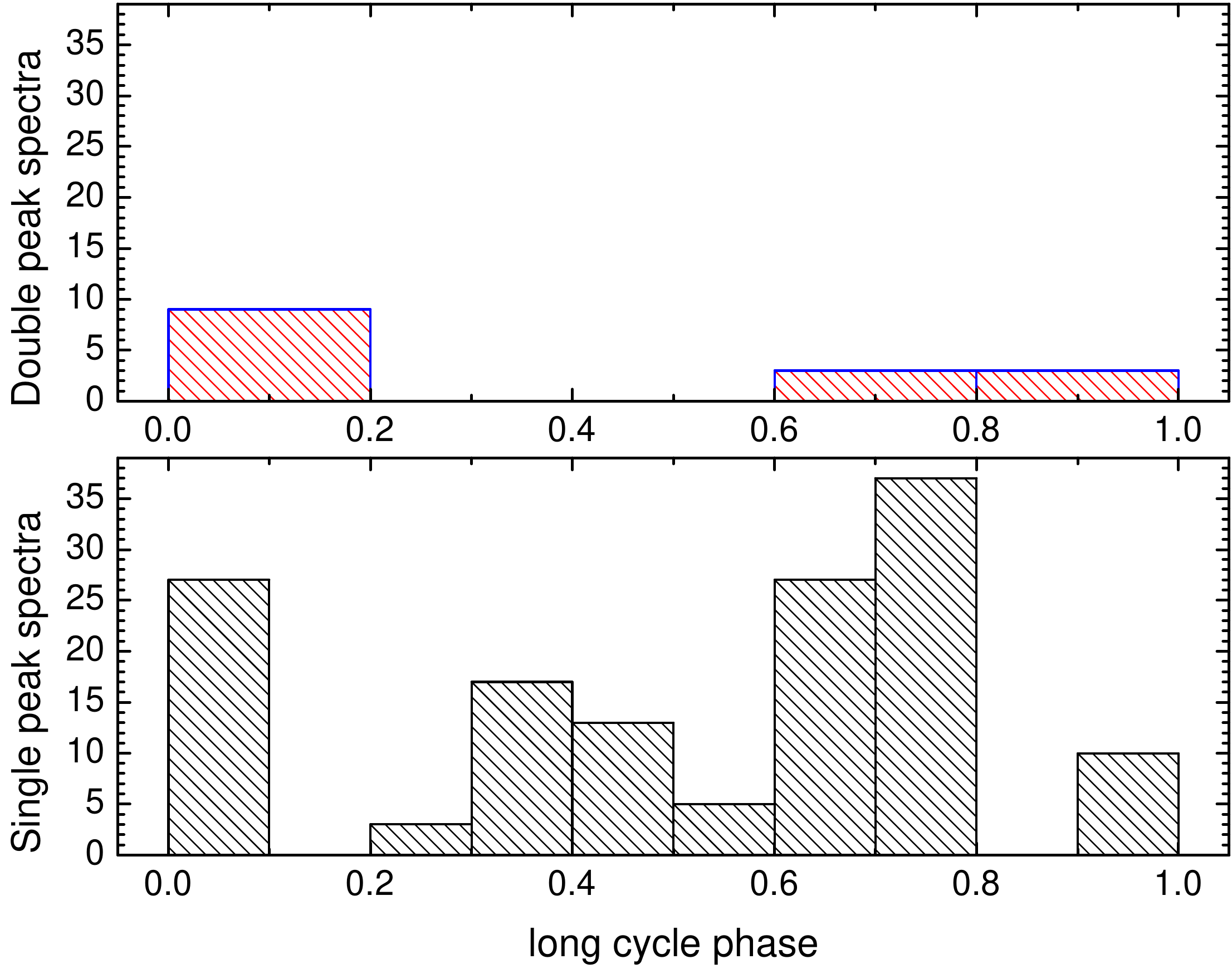}}
\caption{ Left: Radial velocities for the Mg\,II\,4481 emission line. Some illustrative  synthetic RV curves are also shown. Right: Histograms showing the occurrence of single and double peak emission.   }
  \label{y}
\end{figure}

\begin{figure}
\scalebox{1}[1]{\includegraphics[angle=0,width=7cm]{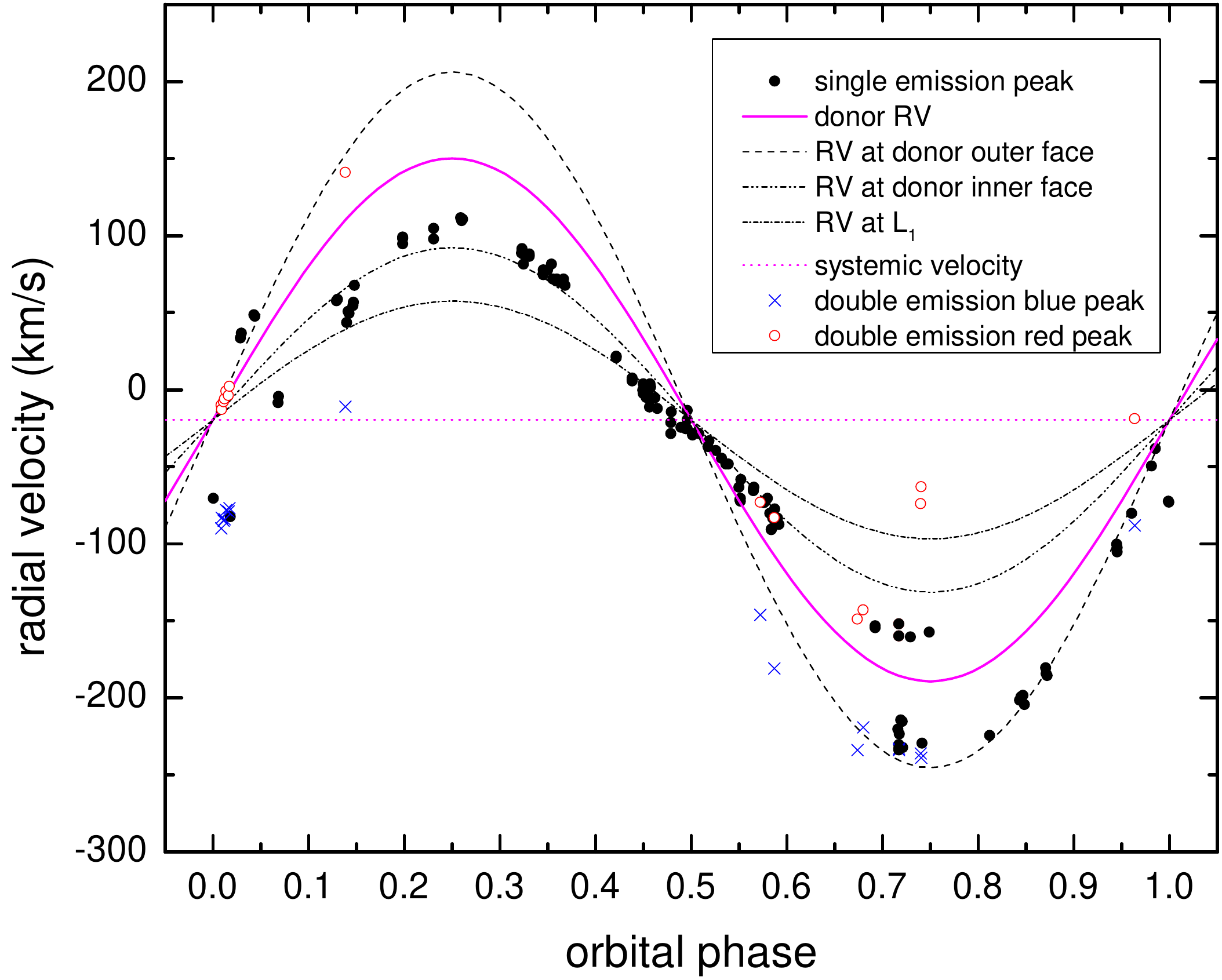}}
\scalebox{1}[1]{\includegraphics[angle=0,width=7cm]{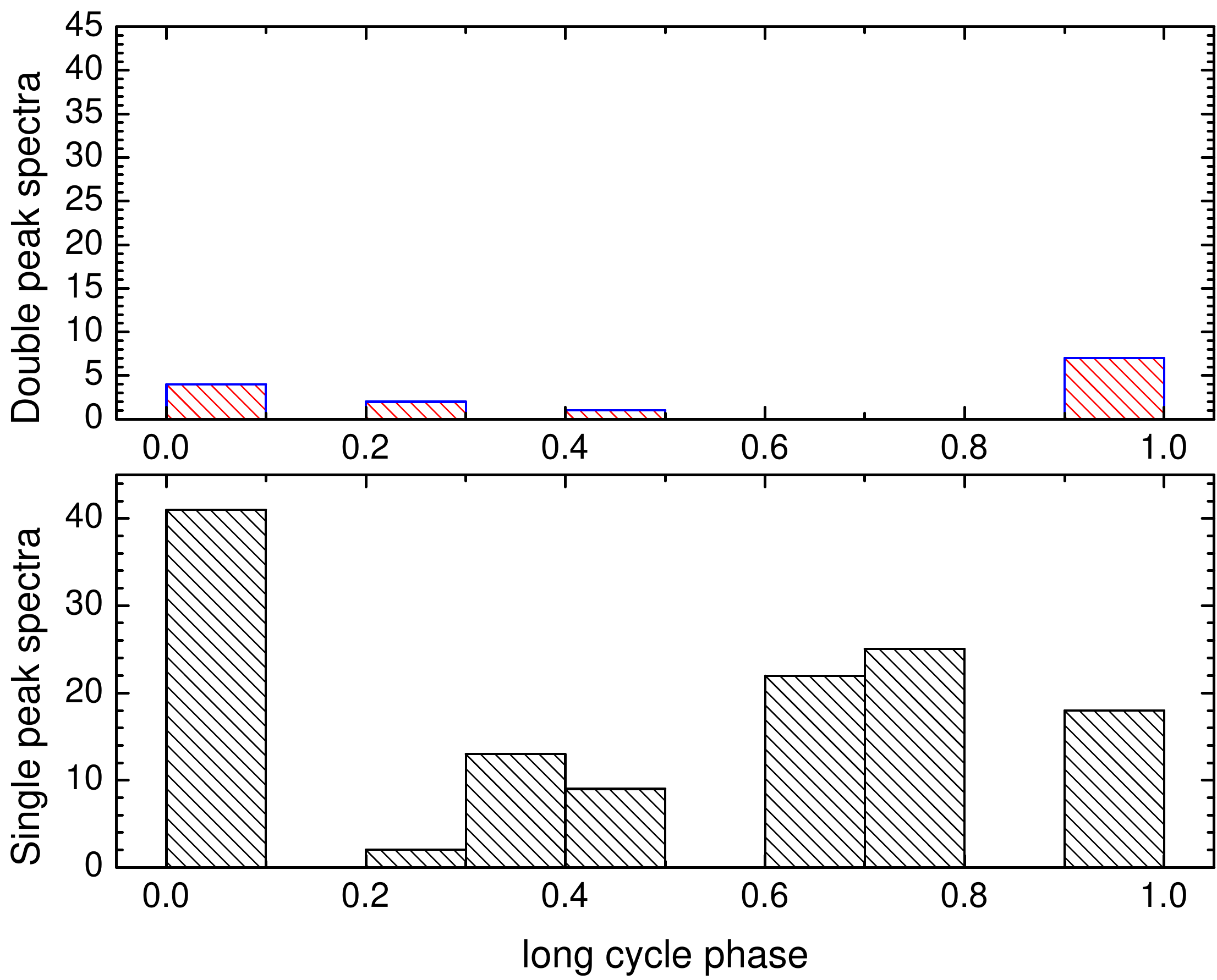}}
\caption{ Left: Radial velocities for the Fe\,I\,4957.6. emission line. Some illustrative  synthetic RV curves are also shown. Right: Histograms showing the occurrence of single and double peak emission.   }
  \label{y}
\end{figure}

\begin{figure}
\scalebox{1}[1]{\includegraphics[angle=0,width=7cm]{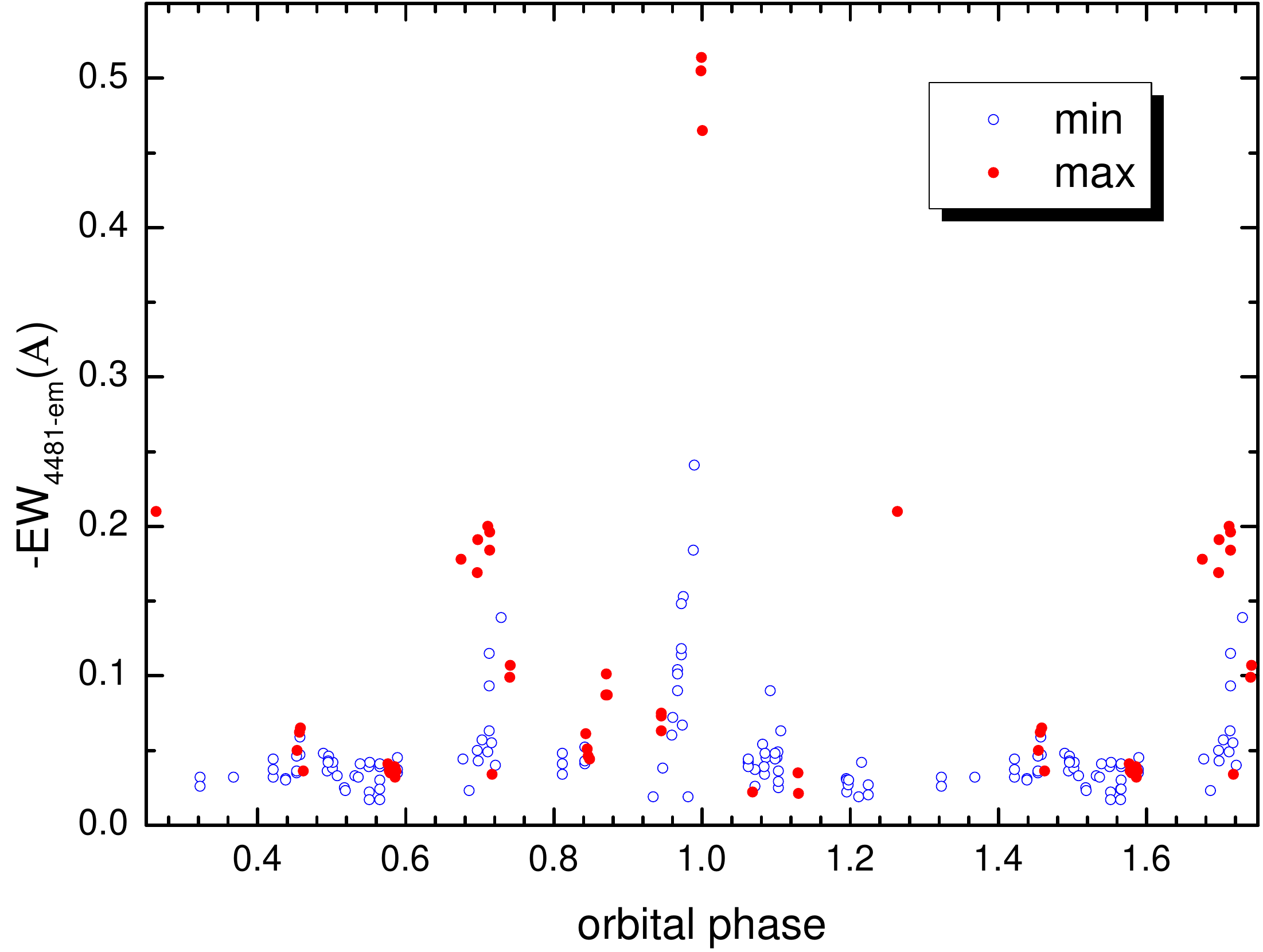}}
\scalebox{1}[1]{\includegraphics[angle=0,width=7cm]{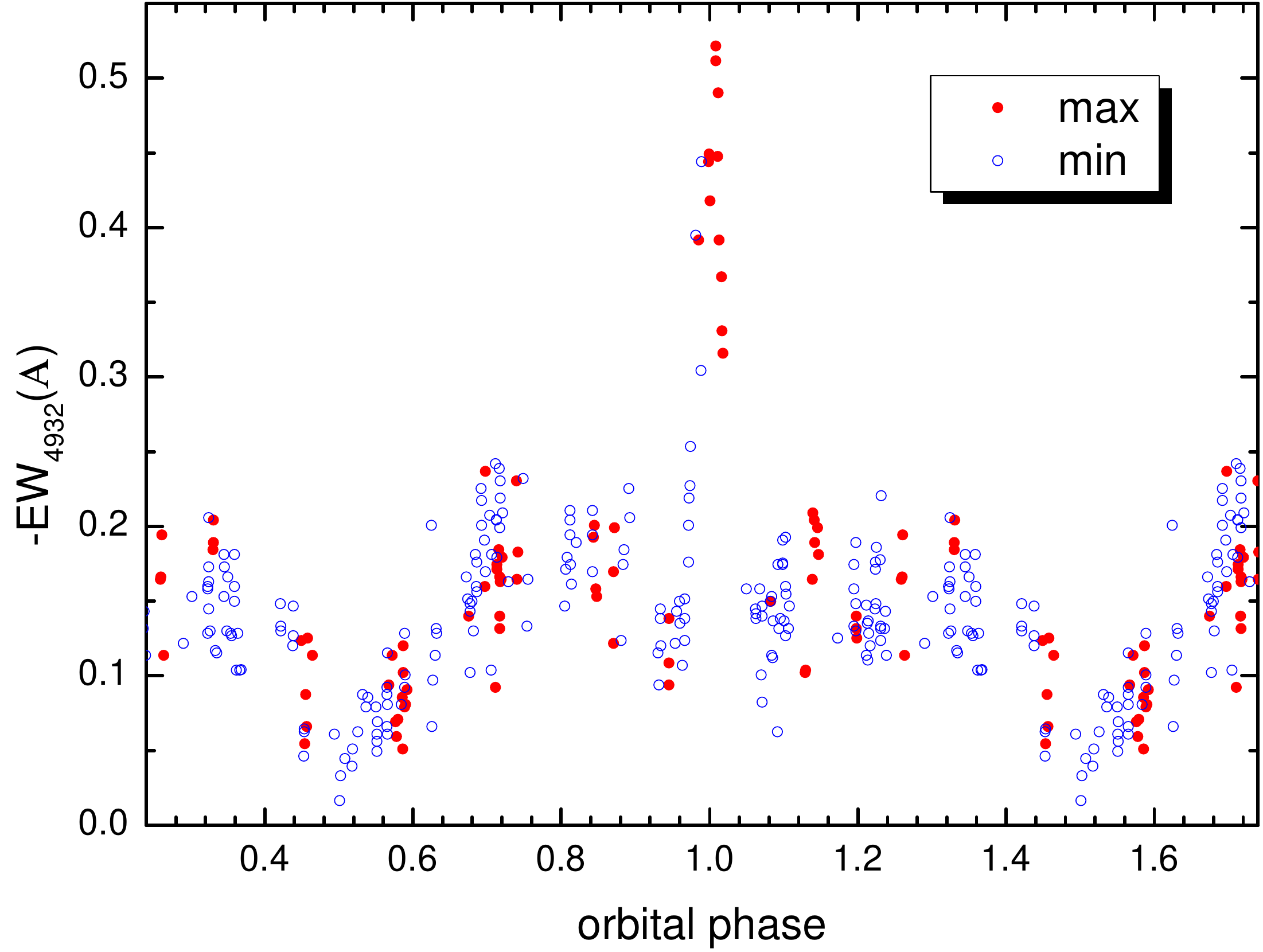}}
\caption{The equivalent width for 0.8 $<$ $\Phi_{l}$ $<$ 0.2 (max) and 0.2 $\leq$ $\Phi_{l}$ $\leq$ 0.8 (min) for Mg\,II\,4481 (left) and C\,I\,4932  (right) emission lines.}
  \label{x}  
\end{figure}

\section{A magnetic dynamo in V\,393\,Scorpii}
\label{S:5}

For the theoretical interpretation of the system, we pursue a fit using the system parameters to the publicly available  \citet{2008A&A...487.1129V}\footnote{http://cdsarc.u-strasbg.fr/viz-bin/Cat?J/A+A/487/1129} models. Through these models, we obtain physical quantities of the donor and gainer stars, such as the convective velocity $v_c$ and the pressure scale height $H_p$. If the long cycle is interpreted to be due to a magnetic dynamo, it can be calculated from the following relation based on \citet{1993ApJ...414L..33S} and \citet{1996ApJ...460..848B}:
\begin{eqnarray*}
P_{long}=D^{\alpha}\times P_{orb},
\end{eqnarray*}
where $D$ is the dynamo number of the donor star and $\alpha$ an index typically between $\frac{1}{3}$ and $\frac{5}{6}$. The dynamo number can be estimated as $D=$Ro$^{-2}$, with Ro the Rossby number. The latter follows form the expression by \citet{2000ApJ...540..436S},
\begin{eqnarray*}
Ro=9\left(\frac{v_{c}}{10\textit{km $s^{-1}$}}\right)\left(\frac{H_{p}}{40\textit{$R_{\odot}$}}\right)^{-1}\left(\frac{\omega}{0.1\omega_{kep}}\right)^{-1}\left(\frac{P_{kep}}{yr}\right),
\end{eqnarray*}
where  $\omega$ is the stellar rotational angular velocity, $\omega_{\rm Kep}$ the Keplerian orbital angular velocity and $P_{\rm Kep}$ is the orbital period of a test particle moving in a Keplerian orbit along the equator of the donor star. These relations have been applied by \citet{2017A&A...602A.109S} to a set of well-studied DPV systems, obtaining a good fit with $\alpha\sim0.31$. More recently, \citet{2018arXiv180307637N} have shown that the model can also successfully reproduce the modulation periods observed in post-common-envelope binaries (PCEBs).

In Fig.\,7 and 8, we show the best-fit binary evolution model by \citet{2008A&A...487.1129V}, focusing on the evolution of mass and radius of the gainer and donor stars, the best fit parameters and the comparison with the observed values (see also \citet{2012MNRAS.421..862M} for procedure and methodology regarding the stellar and orbital parameters). Interesting, we find  also a good agreement between the observed $P_{long}$ and the  prediction of the model.  The log(Plong/Porb)$_{observed}$= 1.516 while log(Plong/Porb)$_{model}$ = 1.576. This means a difference between period ratios between 32.8 and 37.7 i.e. $\sim$ 15 \%, slightly smaller than the value of 20\% obtained by \citep{{2017A&A...602A.109S}} using the observed data and the dynamo model.
We note that the evolutionary stage of the system corresponds to a period  just posterior to  active mass loss, where many of the dynamical parameters are changing as a function of time. This can be seen particularly also in the expected time evolution of the dynamo number and the ratio of long-to-short period given in Fig.\,8, indicating that after an initial rise, the expected long period drops considerably at a stellar age of about 80 million years. It is conceivable that the long periods become observable due to this drop, so that they lie in the range where they can be well-studied and measured. It will be interesting to explore whether a similar behavior can be found in other DPV systems (San Martin et al., in preparation).

\begin{figure}
\scalebox{1}[1]{\includegraphics[angle=0,width=14cm]{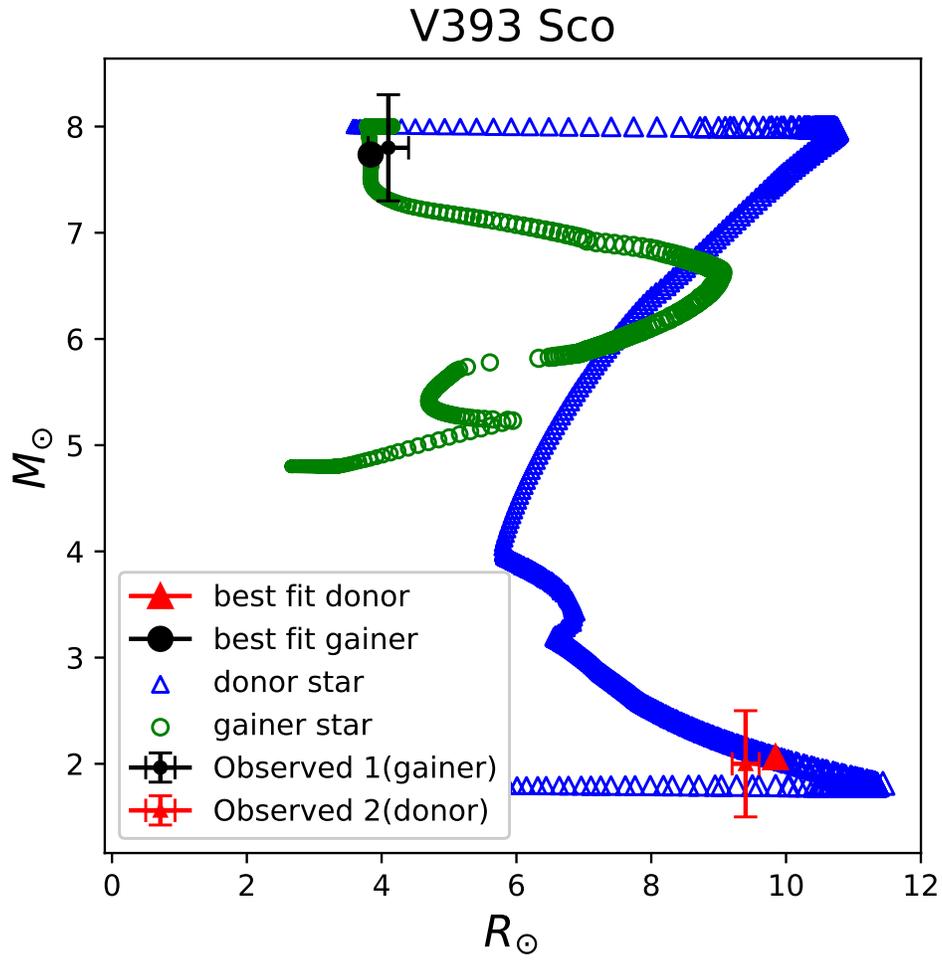}}
\caption{The best  model for V\,393\,Sco according to the \citet{2008A&A...487.1129V} evolutionary tracks and the current system position.}
  \label{x}  
\end{figure}

\begin{figure}
\scalebox{1}[1]{\includegraphics[angle=0,width=7cm]{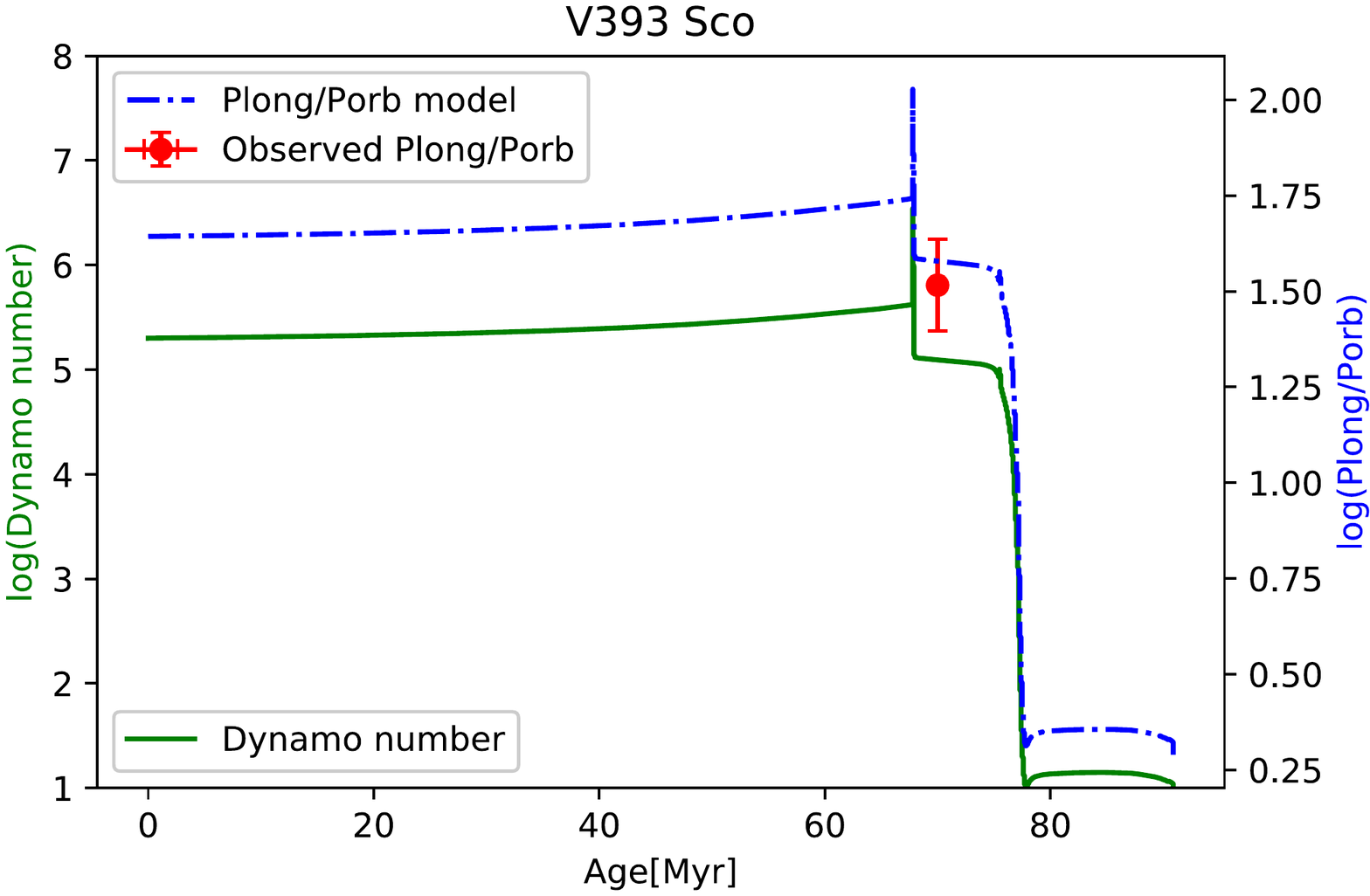}}
\scalebox{1}[1]{\includegraphics[angle=0,width=7cm]{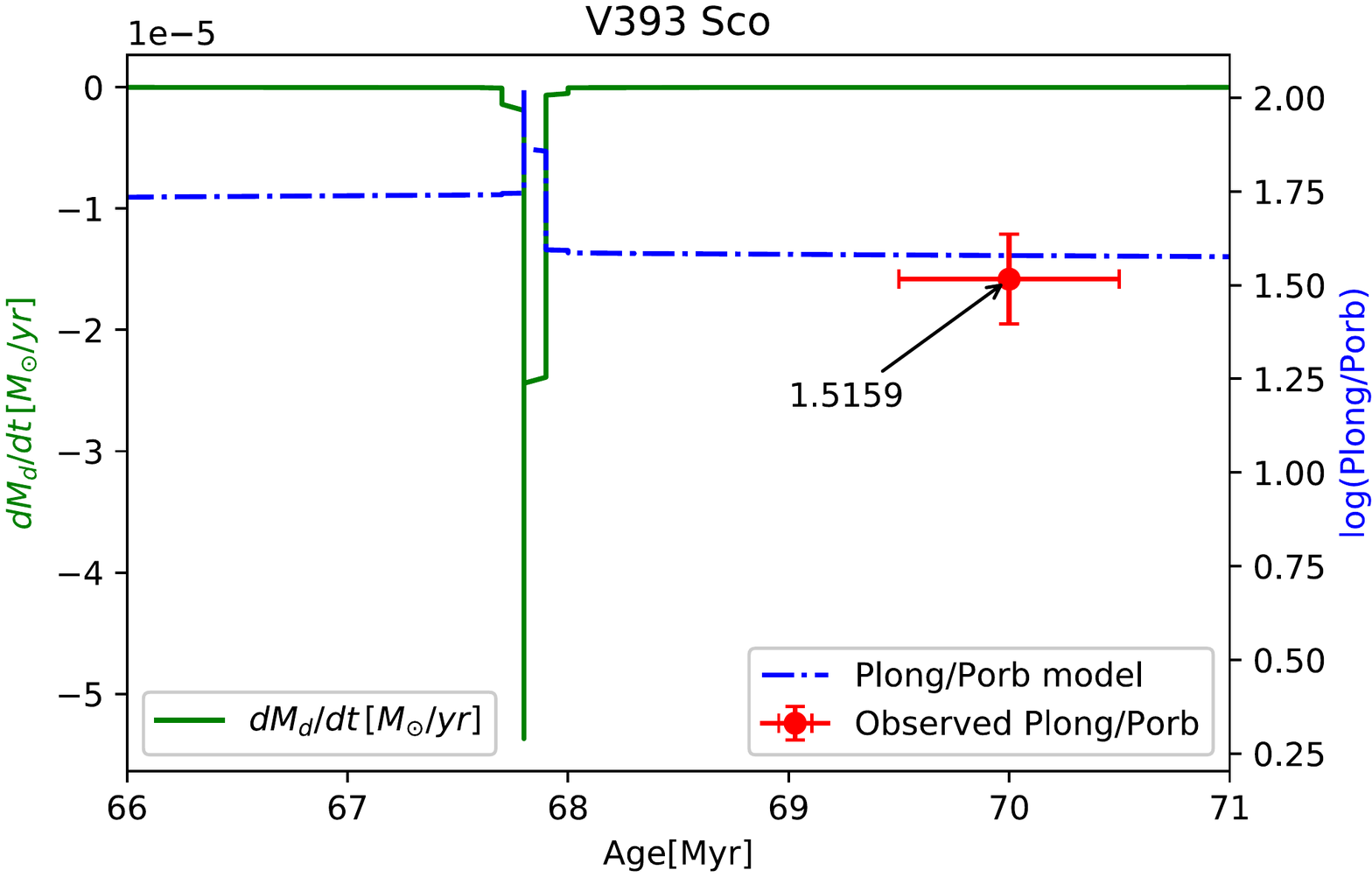}}
\caption{ Left: The evolution of the dynamo number and period ratio for V\,393\,Sco. Right: 
The evolution of the mass transfer rate and period ratio enlarging the epoch of mass transfer.}
  \label{x}  
\end{figure}


\section{Conclusions}

In this paper we report the appearance of  metallic emission lines  in the interacting binary and Double Periodic Variable V\,393\,Sco.
 Some of the Fe\,II and Ti\,II lines are reported for the first time in the literature of this system.
The radial velocity of these features and its variability suggest an origin in a chromospherically active donor star.  Especially interesting are the shifts observed in the radial velocity of some of these lines during the orbital cycle, which might indicate an origin in active regions filling a fraction of the surface of the donor star. The emission of these features
increases during long-cycle maximum, suggesting that the donor is more active at these epochs, eventually increasing the mass transfer onto the gainer.
This should naturally explain the larger Balmer emission observed during high state. The picture emerges of a long cycle reflecting changes in mass transfer which are
modulated by magnetic activity in the donor. In extreme cases, the disc itself might change its radius modulated by variable mass transfer \citep{2018MNRAS.477L..11G}. 
We tested the above assumption applying the \citet{2017A&A...602A.109S} model for the long DPV cycle of V\,393\,Sco obtaining good agreement.
In particular, we observe that the system is found  after a stage of rapid mass transfer and rapid changes in dynamo number; the model predicts a length of the long cycle
very close to the observed one.

\section*{Acknowledgments}

 We acknowledge an anonymous referee for useful comments regarding the first version of this paper.
R.E.M. acknowledges support by VRID-Enlace 216.016.002-1.0 and the BASAL Centro de Astrof{\'{i}}sica y Tecnolog{\'{i}}as Afines (CATA) PFB--06/2007. 
DRGS and RISMP thank for funding via Fondecyt regular (project code 1161247) and from the ''Concurso Proyectos Internacionales de Investigaci\'on, Convocatoria 2015'' (project code PII20150171).






\begin{thebibliography}{}


\bibitem[\protect\citeauthoryear{Baliunas et al.}{1996}]{1996ApJ...460..848B} Baliunas S.~L., Nesme-Ribes E., Sokoloff D., Soon W.~H., 1996, ApJ, 460, 848 

\bibitem[\protect\citeauthoryear{Barr{\'{\i}}a et al.}{2014}]{2014A&A...567A.140B} Barr{\'{\i}}a D., Mennickent R.~E., Graczyk D., Ko{\l}aczkowski Z., 2014, A\&A, 567, A140 


\bibitem[\protect\citeauthoryear{Barr{\'{\i}}a et al.}{2013}]{2013A&A...552A..63B} Barr{\'{\i}}a D., Mennickent R.~E., Schmidtobreick L., Djura{\v s}evi{\'c} G., Ko{\l}aczkowski Z., Michalska G., Vu{\v c}kovi{\'c} M., Niemczura E., 2013, A\&A, 552, A63 

\bibitem[\protect\citeauthoryear{Bolton}{1989}]{1989SSRv...50..311B} Bolton C.~T., 1989, SSRv, 50, 311 


\bibitem[\protect\citeauthoryear{Featherstone et al.}{2009}]{2009ApJ...705.1000F} Featherstone N.~A., Browning M.~K., Brun A.~S., Toomre J., 2009, ApJ, 705, 1000 


\bibitem[\protect\citeauthoryear{Garc{\'e}s L et al.}{2018}]{2018MNRAS.477L..11G} Garc{\'e}s L J., Mennickent R.~E., Djura{\v s}evi{\'c} G., Poleski R., Soszy{\'n}ski I., 2018, MNRAS, 477, L11


\bibitem[\protect\citeauthoryear{Garrido et al.}{2013}]{2013MNRAS.428.1594G} Garrido H.~E., Mennickent R.~E., Djura{\v s}evi{\'c} G., Ko{\l}aczkowski Z., Niemczura E., Mennekens N., 2013, MNRAS, 428, 1594 


\bibitem[\protect\citeauthoryear{Hall}{1989}]{1989SSRv...50..219H} Hall D.~S., 1989, SSRv, 50, 219 


\bibitem[\protect\citeauthoryear{Mennickent}{2017}]{2017SerAJ.194....1M} Mennickent R.~E., 2017, SerAJ, 194, 1 


\bibitem[\protect\citeauthoryear{Mennickent et al.}{2003}]{2003A&A...399L..47M} Mennickent R.~E., Pietrzy{\'n}ski G., Diaz M., Gieren W., 2003, A\&A, 399, L47 



\bibitem[\protect\citeauthoryear{Mennickent et al.}{2008}]{2008MNRAS.389.1605M} Mennickent R.~E., Ko{\l}aczkowski Z., Michalska G., Pietrzy{\'n}ski G., Gallardo R., Cidale L., Granada A., Gieren W., 2008, MNRAS, 389, 1605 

\bibitem[\protect\citeauthoryear{Mennickent et al.}{2010}]{2010MNRAS.405.1947M} Mennickent R.~E., Ko{\l}aczkowski Z., Graczyk D., Ojeda J., 2010, MNRAS, 405, 1947 


\bibitem[\protect\citeauthoryear{Mennickent et al.}{2012a}]{2012MNRAS.421..862M} Mennickent R.~E., Djura{\v s}evi{\'c} G., Ko{\l}aczkowski Z., Michalska G., 2012, MNRAS, 421, 862 

\bibitem[\protect\citeauthoryear{Mennickent et al.}{2012b}]{2012MNRAS.427..607M} Mennickent R.~E., Ko{\l}aczkowski Z., Djurasevic G., Niemczura E., Diaz M., Cur{\'e} M., Araya I., Peters G.~J., 2012, MNRAS, 427, 607 

\bibitem[\protect\citeauthoryear{Mennickent et al.}{2016}]{2016MNRAS.461.1674M} Mennickent R.~E., Zharikov S., Cabezas M., Djura{\v s}evi{\'c} G., 2016, MNRAS, 461, 1674 


\bibitem[\protect\citeauthoryear{Mennickent, Otero, \& Ko{\l}aczkowski}{2016}]{2016MNRAS.455.1728M} Mennickent R.~E., Otero S., Ko{\l}aczkowski Z., 2016, MNRAS, 455, 1728 



\bibitem[\protect\citeauthoryear{Navarrete et al.}{2018}]{2018arXiv180307637N} Navarrete F.~H., Schleicher D.~R.~G., Zamponi J., V{\"o}lschow M., 2018, arXiv, arXiv:1803.07637 


\bibitem[\protect\citeauthoryear{Pawlak et al.}{2013}]{2013AcA....63..323P} Pawlak M., et al., 2013, AcA, 63, 323 


\bibitem[\protect\citeauthoryear{Poleski et al.}{2010}]{2010AcA....60..179P} Poleski R., Soszy{\'n}ski I., Udalski A., Szyma{\'n}ski M.~K., Kubiak M., Pietrzy{\'n}ski G., Wyrzykowski {\L}., Ulaczyk K., 2010, AcA, 60, 179 



\bibitem[\protect\citeauthoryear{Richards et al.}{2014}]{2014ApJ...795..160R} Richards M.~T., Cocking A.~S., Fisher J.~G., Conover M.~J., 2014, ApJ, 795, 160 


\bibitem[\protect\citeauthoryear{Richards \& Albright}{1993}]{1993ApJS...88..199R} Richards M.~T., Albright G.~E., 1993, ApJS, 88, 199 


\bibitem[\protect\citeauthoryear{Schleicher \& Mennickent}{2017}]{2017A&A...602A.109S} Schleicher D.~R.~G., Mennickent R.~E., 2017, A\&A, 602, A109 


\bibitem[\protect\citeauthoryear{Singh, Drake, \& White}{1995}]{1995ApJ...445..840S} Singh K.~P., Drake S.~A., White N.~E., 1995, ApJ, 445, 840 

\bibitem[\protect\citeauthoryear{Soker}{2000}]{2000ApJ...540..436S} Soker N., 2000, ApJ, 540, 436 

\bibitem[\protect\citeauthoryear{Soon, Baliunas, \& Zhang}{1993}]{1993ApJ...414L..33S} Soon W.~H., Baliunas S.~L., Zhang Q., 1993, ApJ, 414, L33 

\bibitem[\protect\citeauthoryear{van Rensbergen et al.}{2008}]{2008A&A...487.1129V} van Rensbergen W., De Greve J.~P., De Loore C., Mennekens N., 2008, A\&A, 487, 1129 



\end{thebibliography}







\end{document}